\documentstyle[prb,preprint,aps]{revtex}
 
\newcommand{\be}{\begin{eqnarray}}
\newcommand{\ee}{\end{eqnarray}}

\begin{document}


\title{\bf Defect Statistics in the Two Dimensional Complex Ginsburg-Landau Model}
\author{Gene F. Mazenko }
\address{The James Franck Institute and the Department of Physics \\
         The University of Chicago \\
         Chicago, Illinois 60637 }
\date{\today}
\maketitle
%
%
\begin{abstract}

The statistical correlations between defects in the two dimensional
complex Ginsburg-Landau model are studied in the defect-coarsening regime.
In particular the defect-velocity probability distribution is
determined and has the same high velocity tail found for the purely
dissipative time-dependent Ginsburg-Landau (TDGL) model.  The spiral arms 
of the defects lead to a very
different behavior for the order parameter correlation function in the 
scaling regime compared to the results for the 
TDGL model.

\end{abstract}
\draft

\section{Introduction}

We study here the statistical properties of 
a collection of point defects generated during the evolution of 
the two-dimensional complex Ginzburg-Landau equation 
(CGLE)\cite{CH,Pismenchap4,AK}.  We 
will be interested in that portion of
the parameter space where the CGLE, driven by random initial
conditions, has a regime of defect coarsening where the density of
defects falls off with a power law in time. 
Our interest here is in the statistical
properties of these defects and ultimately properties of the associated
order parameter driven by the dynamics of the defects.  Initially we will focus on
the velocity distribution of the defects and the spatial
correlations between defects.

The approach developed here is based on the use of a set
topological invariants which are applicable 
to a large set of systems which generate defects as a part of an ordering
process.
In particular one is led to a clean expression
for the velocity of the defect cores in terms of derivatives of the order parameter
field evaluated at the core position.
This approach not only allows one to investigate
equations of motion obeyed by individual defects, but opens up the possibility
of treating the statistical properties of an ensemble of interacting defects.
We have, from previous work\cite{N.8} in the area of 
phase-ordering kinetics\cite{221},
the analog of the Maxwell velocity  distribution 
for a collection of phase-ordering defects. 

In the defect coarsening regime for the CGLE, the defect density, 
$\bar{n}(t)$,
scales as $L^{-2}(t)$ where $L(t)$ is a  characteristic length
which grows with $t$, 
and $t$ is the time of the evolution of the
system starting with
random initial conditions.  In these
circumstances, as shown in detail below, the 
defect velocity probability distribution is given, as in the
purely dissipative time-dependent Ginzburg-Landau (TDGL) case, by
\be
P({\bf V})=\Gamma (\frac{3}{2})
\Biggl(\frac{1}{\pi \bar{v}^{2}}\Biggr) 
\Biggl( 1+{\bf V}^{2}/\bar{v}^{2}\Biggr)^{-2}
\label{eq:a1}
~~~,
\ee
where the characteristic velocity $\bar{v}\approx L^{-1}$ is given explicitly
below.
Similarly the defect-defect equal-time correlation function
has the same form (see below) as found for the TDGL case.  In
the case of the correlations between defect densities
at different times, we find some rather weak deviations from the
results in the TDGL case.

These results for the statistical properties of the defects, inspire
one, using ideas which have been successful for treating the
TDGL case, to look at the order parameter correlations.  In
this case we find 
results quite different from the TDGL case.  This is due both
to the spiral arms and processional motion characteristic of
defects in the CGLE. The spiral arms render order parameter correlations shorter
in range, compared to the TDGL case, and the order parameter correlation function
shows the behavior $\approx \bar{n}^{3}(t)W(r/L(t))$.
The precessional effects are predicted to be prominent in the two-time
order parameter correlation function.

\section{Background}

The complex Ginzburg-Landau equation can be written\cite{CH} in the form
\be
\partial_{t} \psi =b \nabla^{2}\psi +\left(1-u|\psi |^{2}\right)\psi
\label{eq:4}
\ee
where $\psi$ is a complex field,
and  $b$ and $u$ are complex parameters.
For the appropriate set of parameters (choice of $b$ and $u$)
we find on
quenching from an initially disordered state, that the CGLE
generates a set of coarsening point defects.
The characteristic distance between the defects increases with
time due to the annihilation process between defects and antidefects\cite{glass}.

For $b$ and $u$ real  Eq.(\ref{eq:4})
reduces to the dissipative TDGL equation
which is the most widely studied model for phase ordering\cite{221}.
If we set $b=u=i\eta$ in Eq.(\ref{eq:4}) and  take  $\eta$ large,
we find, and after a simple gauge
transformation, Eq.(\ref{eq:4}) reduces to 
\be
-i\partial_{t} \psi = \nabla^{2}\psi +(1-|\psi |^{2})\psi
~~~.
\ee
This equation, the nonlinear Schroedinger equation  (NLSE)\cite{Gross},
gives a highly idealized description of the low temperature properties of a
neutral superfluid.
Unlike the TDGL
system, the NLSE supports several conserved quantities.  In particular the quantity
$\int d^{d}r ~|\psi ({\bf r},t)|^{2}$ does not change with time.  This model supports
the same defects as the TDGL model, but the dynamics of the defects are quite
different.  Two oppositely charged vortices in the TDGL model
move along the line connecting them toward annihilation.  In the NLSE the same
two vortices move at right angles to the line connecting them.  

\section{Defect Charged Density and Velocity Fields}

The approach developed here allows for a direct connection between a set of field
equations, like the complex Ginzburg-Landau equation satisfied by an order
parameter field, and the
equations of motion  of the cores of a set of defects.
It has only recently  been understood, as discussed below, that these
expressions for the defect velocity  reduce to the same  form
as found in pattern forming studies using very different
arguments.

The approach developed here is motivated by addressing
the question:
What is the probability of finding a defect a distance $r$ from an
anti-defect? In work on phase ordering kinetics we\cite{cs14} developed
methods which are convenient for handling such questions.
A motivating factor was the realization that
in treating statistical properties
of defects one does not want to work with formal structures
which require  an explicit
treatment in terms of the defect positions.  This leads to problems  
of specification of initial conditions.
Instead
we looked for  a way of implicitly finding the positions of the defects
using the order parameter field $\psi$ itself.

Let us consider the case of two dimensions where we have point defects.
The case of line defects can also be treated\cite{ctg,string}  using these ideas but
will be discussed elsewhere.
The basic idea is that the positions of
defects are located by the zeros\cite{Rice,HL66,HAL81,cs14} of the 
order parameter field
$\psi$.  Suppose, instead of the positions ${\bf r}_{i}(t)$ we want
to write our description in terms of the zeros of $\psi({\bf r},t)$.
It is not difficult to see that the defect charged density has the
two representations
\be
\rho ({\bf r},t)=\delta(\vec{ \psi}({\bf r},t)){\cal D}({\bf r},t)
=\sum_{i=1}^{N}q_{i}\delta ({\bf r}-{\bf r}_{i}(t))
\label{eq:3.1}
\ee
where
$q_{i}={\cal D}({\bf r}_{i})/|{\cal D}({\bf r}_{i})|=\pm 1$,
and ${\cal D}({\bf r})$ is the Jacobian
associated with 
the change of variables
from the set of defect positions to the field $\psi$:
\be
{\cal D}=\frac{i}{2!}\epsilon_{\mu_{1}\mu_{2}}
\nabla_{\mu_{1}}\psi
\nabla_{\mu_{2}}\psi^{*}
\label{eq:det}
\ee
where we sum over the $\mu_{i}$,  $\epsilon_{\mu_{1}\mu_{2}}$ is the
$2$-dimensional anti-symmetric tensor and
summation over repeated indices is implied. 
For later reference,
the unsigned defect density is given by
$n ({\bf r},t)= |\rho ({\bf r},t)|$.

For systems where only unit charges are present,
$\rho$ is the topological charge density.  Notice
that $q_{i}$ is well defined even for system like
classical fluids where the circulation associated
with a defect is not quantized.

The dynamical
implications of this approach are simple.  If
indeed topological charge is conserved then we
would expect the charge density to obey a continuity equation. 
It was shown in
Ref.(\onlinecite{N.8}) that $\rho$ satisfies
a continuity equation of the form
\be
\partial _{t}\rho =-\vec{\nabla}\cdot \left(\rho {\bf v}\right)
\label{eq:10}
\ee
where the defect velocity field ${\bf v}$  is given explicitly by
\be
{\cal D} v_{\alpha}=-\frac{i}{2}\sum_{\beta}\epsilon_{\alpha\beta}
\left(\dot{\psi}\nabla_{\beta}\psi^{*}
-\dot{\psi}^{*}\nabla_{\beta}\psi\right)
~~~.
\label{eq:12}
\ee
where  ${\cal D}$ is defined by Eq.(\ref{eq:det}) and we must remember
that ${\bf v}$ is multiplied by the defect core locating $\delta$-function
in $\rho$ in Eq.(\ref{eq:10}).
Eq.(\ref{eq:12}) gives one an explicit expression for the defect velocity 
field  expressed
in terms of derivatives of the order parameter.
This expression for the defect velocity seems  
to be very general.  Notice that we have not specified the form
of the equation of motion for the order parameter only that the order
parameter be complex and $d=2$.
For the CGLE our expression for the defect velocity
reduces to
\be
{\cal D} v_{\alpha}=-\frac{i}{2}\sum_{\beta}\epsilon_{\alpha\beta}
\left(b\nabla^{2}\psi\nabla_{\beta}\psi^{*}
-b^{*}\nabla^{2}\psi^{*}\nabla_{\beta}\psi\right)
~~~.
\label{eq:36}
\ee
Does this expression for the velocity agree with our expectations for known cases?  
Let us assume that we have a defect of charge $m$ at the origin of our two
dimensional system and write the order parameter in the form:
$\psi =Re^{i\theta}$, 
$R=r^{|m|}e^{w}$ and $\theta = m\phi +\theta_{B}$,
where again $r$ and $\phi$ are the cylindrical coordinates relative to the
core at the origin.
It is then a straightforward bit of calculus to show that the velocity given by 
Eq.(\ref{eq:36}) reduces to
\be
v_{\alpha}=2b''\left(\nabla_{\alpha} \theta_{B}
+\frac{m}{|m|}\sum_{\beta}\epsilon_{\alpha\beta}\nabla_{\beta}w\right)
-2b'\left(\nabla_{\alpha} w
-\frac{m}{|m|}\sum_{\beta}\epsilon_{\alpha\beta}\nabla_{\beta}\theta_{B}\right)
~~~.
\label{eq:40}
\ee
If we ignore the contributions due to the variation in the amplitude, $w$,
Eq.(\ref{eq:40}) reduces\cite{gauge} to
$v_{\alpha}=2b''\nabla_{\alpha} \theta_{B}
+b'\frac{m}{|m|}\sum_{\beta}\epsilon_{\alpha\beta}\nabla_{\beta}\theta_{B}$.
The first term is the only contribution in the NLSE case and states that
a vortex moves with the local superfluid velocity\cite{Fetter66}.  The second
term is the Peach-Koehler\cite{Peach} term first found in this context
by Kawasaki\cite{Kawasaki84}.  These are the results from the phase-field
approach and leads, for example, to the same type of interaction between 
two vortices as found in fluids.
The velocity of
a single isolated vortex is zero.  For a set of two isolated vortices
one has the expected behavior for the TDGL and NLSE cases. 

For our purposes here the more
important point is to consider the work of T$\ddot{o}$rnkvist
and Schr$\ddot{o}$der\cite{Tornkvist}.  Using methods of differential geometry,
they looked at the derivation of the form of the velocity of a defect in
the case of the CGLE.  They comment,
"The evolution of a system with (spiral) vortices may be described in 
terms of the defects, or filaments, along with values of the fields"
$R$ and $\theta$ "at positions away from the defects of filaments.  Such
a separation into collective coordinates and field variables is non-trivial,
and the present work comprises the first exact treatment of this kind for a
dissipative system".  The final equation they obtain, in our notation here
and for two-dimensional systems, is precisely given by Eq.(\ref{eq:40}).
Thus the velocity given by Eq.(\ref{eq:40}) reproduces the most sophisticated
results obtained using other methods.

\section{Auxiliary Field Method}

\subsection{Overview}

How can we use these expressions for $\rho$ and ${\bf v}(\psi)$ to compute
the measureable statistical properties of an evolving CGLE system?  We will
use a generalization of an approximate method which has led to good
results for the TDGL case. The basic idea is to assume that there is 
mapping from the order parameter field onto an auxiliary field $m$ which
shares the same zeros as $\psi$ in space.  In particularly we require
$\rho [\psi]=\rho [m]$ and
${\bf v}[\psi]={\bf v}[m]$ where again we use the result that the velocity is
 multiplied by a defect-zero-
finding $\delta$-function.  These requirements are not very constraining since
they only require that $\psi$ be proportional to $m$ for small $m$ with corrections
that are cubic in $m$.  
It has been convenient to think of ${\bf m}({\bf x})$ as a 2-vector whose magnitude gives the
distance from ${\bf x}$ to a defect core.  Thus, as discussed in more detail 
below, near the core we can take $Re ~ \psi =m_{x}$ and $Im  ~\psi =m_{y}$.

The main assumption\cite{ctg} in the theory is that the field
$m$ is gaussian and the variance in $m$ is determined by requiring that
the defect charge density continuity equation be satisfied on average:
\be
\frac{\partial G_{\rho\rho}(12)}{\partial t_{1}}
=-\vec{\nabla}_{1}\cdot <\rho (1){\bf v}(1)\rho (2)> \equiv G_{J\rho}(12)
~~
\label{eq:13}
\ee
where 
\be
G_{\rho\rho}(12)=<\rho (1) \rho (2)>
\label{eq:15}
\ee
and $\rho (1)=\rho (t_{1},{\bf x}_{1})$.
With these assumptions and assumptions about the initial conditions, one
can work out all of the statistical properties of the defects including
$G_{\rho\rho}(12)$ and the defect velocity
probability distribution function defined by
\be
\bar{n} P({\bf V})\equiv
\langle |\rho | \delta ({\bf V}-{\bf v}[\psi])\rangle ~~~,
\label{eq:42}
~~~.
\ee
Corrections to this gaussian approximation can be investigated using methods
of the type developed in Refs.\onlinecite{EXP,EXP2}.

The procedure then  is to first compute $G_{\rho\rho}(12)$ and 
$G_{J\rho}(12)$
assuming that $m$ is a gaussian field.  This will give 
$G_{\rho\rho}(12)$ and $G_{J\rho}(12)$ as functions of the
auxiliary field correlation function
\be
C_{\alpha\beta} (12) = <m_{\alpha}(1) m_{\beta}(2)>
\ee
where $\alpha$ and $\beta$ take on the values $x$ and $y$.
Inserting these results for $G_{\rho\rho}(12)$ and $G_{J\rho}(12)$
back into Eq.(\ref{eq:13}) gives an equation for $C(12)$.  It
will turn out that this equation for $C(12)$ can be solved analytically.
This result can then be fed back into the result for $G_{\rho\rho}(12)$ to obtain an
explicit expression for the defect density correlation function. As part of
this calculation we obtain the average defect density $\bar{n}=<|\rho |>$.
Finally we can carry out the average over the gaussian variable $m$ to obtain
$P({\bf V})$ as a function of $C(12)$, and in turn obtain an explicit
expression for $P({\bf V})$.

\subsection{Expressing $G_{\rho\rho}$ in terms of $C$}

The defect density in the defect-defect correlation function
defined by Eq.(\ref{eq:15}) can be  written explicitly in terms of the
gaussian auxiliary field ${\bf m}$ in the form
\be
\rho (1)=\frac{1}{2}\epsilon_{\mu_{1}\mu_{2}}\epsilon_{\nu_{1}\nu_{2}}
\nabla_{\mu_{1}}m_{\nu_{1}}(1)\nabla_{\mu_{2}}m_{\nu_{2}}(1)
\delta ({\bf m}(1))
\ee
and we sum over all the indices $\nu$ and $\mu$.  
In the isotropic case, worked out previously, the evaluation of
$G_{\rho\rho}$ for the $n$-vector model for the general case of $n=d$
was facilitated by the decomposition of the average for $G_{\rho\rho}$ into
a product of averages corresponding to each component.  This decomposition is
not possible here because the complex coefficients in the CGLE couple the
components of the
order parameter as the system evolves.  We need a more general approach.
This more general approach involves using the general identity valid for
Gaussian fields:
\be
<m_{\nu}(1)F[{\bf m}]>=\sum_{\nu '}\int ~dt_{2}d^{2}x_{2}~  C_{\nu\nu'}(12)
<\frac{\delta }{\delta m_{\nu'}(2)}F[{\bf m}]>
~~~.
\ee
Using this result for all of the fields
in $G_{\rho\rho}$ acted upon by a gradient in Eq.(\ref{eq:15}),
one can bring all of the gradients
outside  the average.  This generates many terms which are products of
the matrix $C$ and averages proportional to the quantities
\be
{\cal G}(12)=<\delta ({\bf m}(1))\delta ({\bf m}(2))>
~~~,
\ee
and
\be
{\cal G}_{\nu_{1}\nu_{2}}(12)
=<\frac{\partial}{\partial m_{\nu_{1}}(1)}\delta ({\bf m}(1))
\frac{\partial}{\partial m_{\nu_{2}}(2)}\delta ({\bf m}(2))>
~~~,
\ee
and similar higher-order derivatives of the $\delta$-functions
which do not contribute to the final result.

A key assumption in the evaluation of $G_{\rho\rho}$ is that the system is
isotropic in space and we can write:
\be
C_{\nu\nu'}(12)=C_{\nu\nu'}(r,t_{1}t_{2})
~~~,
\ee
\be
\nabla_{\mu}^{(1)}C_{\nu\nu'}(12)=C^{\prime}_{\nu\nu'}(12)\hat{r}_{\mu}
~~~,
\ee
and
\be
\nabla_{\mu}^{(1)}\nabla_{\mu'}^{(2)}C_{\nu\nu'}(12)
=-\left[C^{L}_{\nu\nu'}(12)-C^{T}_{\nu\nu'}(12)\right]
\hat{r}_{\mu}\hat{r}_{\mu'}-C^{T}_{\nu\nu'}(12)\delta_{\mu\mu'}
\ee
where ${\bf r}={\bf x}_{1}-{\bf x}_{2}$, and
\be
C^{L}_{\nu\nu'}(12)\equiv C^{\prime\prime}_{\nu\nu'}(12)
\ee
\be
C^{T}_{\nu\nu'}(12)\equiv \frac{1}{r}C^{\prime}_{\nu\nu'}(12)
\ee
and the primes in the superscripts indicate derivatives with respect to $r$.  
Using these results one can then carry our the sums over the spatial
coordinate labels, the $\mu$'s in $G_{\rho\rho}$
to obtain:
\be
G_{\rho\rho}(12)=G^{(1)}_{\rho\rho}(12)+G^{(2)}_{\rho\rho}(12)
~~~,
\ee
where
\be
G^{(1)}_{\rho\rho}(12)={\cal G}(12)\epsilon_{\nu_{1}\nu_{2}}
\epsilon_{\nu^{\prime}_{1}\nu^{\prime}_{2}}
C^{L}_{\nu_{1}\nu^{\prime}_{1}}(12)C^{T}_{\nu_{2}\nu^{\prime}_{2}}(12)
\ee
and
\be
G^{(2)}_{\rho\rho}(12)=-\epsilon_{\nu_{1}\nu_{2}}
\epsilon_{\nu^{\prime}_{1}\nu^{\prime}_{2}}
{\cal G}_{\sigma_{2}\sigma_{1}}(12)
C^{\prime}_{\nu_{1}\sigma_{1}}(12)C^{\prime}_{\sigma_{2}\nu^{\prime}_{2}}(12)
C^{T}_{\nu_{2}\nu^{\prime}_{1}}(12)
~~~.
\ee
It is easy to evaluate, using the results from Appendix A, the
remaining averages over the auxiliary field:
\be
{\cal G}_{\sigma_{2}\sigma_{1}}(12)=D^{2}
C_{\sigma_{2}\sigma_{1}}(12){\cal G}(12)
\ee
\be
{\cal G}(12)=\frac{D^{2}}{(2\pi)^{2}}
~~~
\ee
where $ D$ is defined by Eq.(\ref{eq:A10}).
Expressing $C_{\nu\nu'}(12)$ in terms of $C_{0}(12)$ and $\Delta (12)$,
as given by Eq.(\ref{eq:A2}),
and doing the sums over the $\nu$'s, we find after some rearrangement,
the result for the defect density correlation function:
\be
G_{\rho\rho}(12)=\frac{1}{(2\pi)^{2}}
\frac{1}{r}\frac{d}{dr}\left(Q\gamma_{T}^{2}\right)
\label{eq:29}
\ee
where
\be
\gamma_{T}^{2}=(1-f_{T}^{2})^{-1}
\label{eq:30}
\ee
\be
f_{T}=\sqrt{f_{0}^{2}+\Delta_{0}^{2}}
\label{eq:31}
\ee
\be
f_{0}=\frac{C_{0}}{\sqrt{S_{0}(1)S_{0}(2)}}
\label{eq:32}
\ee
\be
\Delta_{0}=\frac{\Delta}{\sqrt{S_{0}(1)S_{0}(2)}}
\label{eq:33}
\ee
where $S_{0}(i)=C_{0}(ii)$ and
\be
Q=\left(f_{0}'\right)^{2}+\left(\Delta_{0}'\right)^{2}
~~~.
\ee
We still need to determine the auxiliary field correlation
functions $C_{0}$ and $\Delta$.
It is easy to see that the result given by Eq.(\ref{eq:29}), 
in the isotropic limit where $\Delta =0$,
reduces to result first reported by Halperin\cite{HAL81}
\be
G_{\rho\rho}(12)=\frac{1}{r}\frac{d}{dr}\left(h^{2}\right)
\ee
where
\be
h=\frac{\gamma_{T}f_{0}'}{2\pi}
~~~.
\ee

\subsection{Satisfying Conservation of Topological Charge}

The calculation of the current contribution of $G_{J\rho}$ on the right-hand
side of Eq.(\ref{eq:13}) is much the same as for $G_{\rho\rho}$ except for
terms which involve the  on-site correlation function
\be
S^{(2)}(1)=\frac{1}{2}<\left(\nabla{\bf m}(1)\right)^{2}>
=-\left(\nabla^{2}C_{0}(r,t_{1}t_{1})\right)_{r=0} 
~~~.
\ee
$G_{J\rho}$ is also proportional to the factor
$\frac{1}{(2\pi)^{2}}\frac{1}{r}\frac{d}{dr}$ and, after performing an
integration over $r$, we obtain the  averaged conservation law, given by
Eq.(\ref{eq:13}), can be rewritten as
\be
\frac{\partial}{\partial t_{1}}\left(Q\gamma_{T}^{2}\right)
=2b'M+2b''N
\label{eq:38}
\ee
where
\be
M=\gamma_{T}^{4}Q\left(\omega_{0}(1)+f_{0}\nabla^{2}f_{0}
+\Delta_{0}\nabla^{2}\Delta_{0}\right)+\gamma_{T}^{2}
\left(f_{0}'\nabla^{2}f_{0}'+\Delta_{0}'\nabla^{2}\Delta_{0}'\right)
\ee
and
\be
N=\gamma_{T}^{4}Q\left(f_{0}\nabla^{2}\Delta_{0}-\Delta_{0}\nabla^{2}f_{0}\right)
+\gamma_{T}^{2}
\left(f_{0}'\nabla^{2}\Delta_{0}'+\Delta_{0}'\nabla^{2}f_{0}'\right)
\ee
where we have introduced the time-dependent quantity
\be
\omega_{0}(1)=\frac{S^{(2)}(1)}{S_{0}(1)}
=-\left(\nabla^{2}f_{0}(r,t_{1}t_{1})\right)_{r=0}
~~~.
\label{eq:41}
\ee
Eq.(\ref{eq:38}) looks very complicated but simplifies if we replace
$f_{0}$ and $\Delta_{0}$ with
\be
f_{0}=f_{T}~cos~\Omega
\ee
and
\be
\Delta_{0}=f_{T}~sin~\Omega
~~~.
\ee
Then Eq.(\ref{eq:38}) can be rewritten as
\be
\gamma_{T}f_{T}'\left[\gamma_{T}\left(2\dot{f}_{T}-R\right)\right]'
+\gamma_{T}^{4}f_{T}\left(\Omega'\right)^{2}
\left(2\dot{f}_{T}-R\right)
+\gamma_{T}^{2}\Omega'\left[2f_{T}^{2}\dot{\Omega}'
-f_{T}'S+f_{T}S'\right]=0
\label{eq:WWW}
\ee
where
\be
R=2b'\omega_{0}(1)f_{T}+2b'A+2b''B
\ee
\be
S=-2b'B+2b''A
\ee
and
\be
A=\nabla^{2}f_{T}-f_{T}(\nabla\Omega )^{2}
\ee
\be
B=2\nabla f_{T}\cdot \nabla \Omega +f_{T}\nabla^{2}\Omega
~~~.
\ee
A solution to Eq.(\ref{eq:WWW}) is given by 
\be
2\dot{f}_{T}=R
\label{eq:49}
\ee
and
\be
2f_{T}^{2}\dot{\Omega}'=f_{T}'S-f_{T}S'
~~~.
\ee
This last equation can be reduced to
\be
2f_{T}\dot{\Omega}=-S
\label{eq:51}
\ee
The set of coupled equation given by Eqs.(\ref{eq:49}) and (\ref{eq:51}) 
are equivalent
to the equations for $f_{0}$ and $\Delta_{0}$ given by
\be
\dot{f}_{0}=b'\left(\omega_{0}(1)+\nabla^{2}\right)f_{0}
+b''\nabla^{2}\Delta_{0}
\label{eq:52}
\ee
and
\be
\dot{\Delta}_{0}=b'\left(\omega_{0}(1)+\nabla^{2}\right)\Delta_{0}
-b''\nabla^{2}f_{0}
~~~.
\label{eq:53}
\ee
This is the set of equations which must be  solved self-consistently to obtain
the unknown quantities $f_{0}$ and $\Delta_{0}$ and $\omega (1)$.

\subsection{Auxiliary Field Correlation Function}

Equations (\ref{eq:52}) and (\ref{eq:53}) are reduced to a set of
differential equations in time if we Fourier transform in space
and  put in the time labels explicitly:
\be
\frac{\partial}{\partial t_{1}}f_{0}({\bf q},t_{1}t_{2})
=\alpha (q,t_{1})f_{0}({\bf q},t_{1}t_{2})-\beta_{q}\Delta_{0}({\bf q},t_{1}t_{2})
\label{eq:54}
\ee
\be
\frac{\partial}{\partial t_{1}}\Delta_{0}({\bf q},t_{1}t_{2})
=\alpha (q,t_{1})\Delta_{0}({\bf q},t_{1}t_{2})+\beta_{q}f_{0}({\bf q},t_{1}t_{2})
\label{eq:55}
\ee
where
\be
\alpha (q,t_{1})=b'\left(\omega_{0}(t_{1})-q^{2}\right)
\label{eq:56}
\ee
and
\be
\beta_{q}=b'' q^{2}
\label{eq:57}
~~~.
\ee
Equations (\ref{eq:54}) and (\ref{eq:55}) need to be solved together with the
symmetry condition
\be
f({\bf q},t_{1},t_{2})=f_{0}({\bf q},t_{1},t_{2})+i\Delta_{0}({\bf q},t_{1},t_{2})
=f^{*}(-{\bf q},t_{2},t_{1})
\label{eq:58}
\ee
and the initial condition
\be
f({\bf q},t_{0},t_{0})=2\pi\ell^{2}e^{-\frac{1}{2}(q\ell )^{2}}\equiv g(q)
\label{eq:59}
~~~.
\ee
This particular choice of initial conditions, corresponding to an initial
correlation length $\ell$, is very convenient since all integrals can
be carried out analytically for all times.  Finally we must remember
the normalization
which follows from the definition of $f(12)$ given by Eqs.(\ref{eq:32}) 
and (\ref{eq:33}): 
\be
f(11)=\int ~\frac{d^{2}q}{(2\pi)^{2}} f({\bf q},t_{1},t_{1}) =1
\label{eq:60}
~~~.
\ee

It is not difficult to construct the appropriate solution given by
\be
f({\bf q},t_{1},t_{2})=R(t_{1},t_{0})R(t_{2},t_{0})g(q)
e^{-b'q^{2}(t_{1}+t_{2}-2t_{0})}
e^{i\beta_{q}(t_{1}-t_{2})}
\label{eq:61}
\ee
where
\be
R(t_{1},t_{0})=
e^{b'\int_{t_{0}}^{t_{1}}~d\tau \omega_{0}(\tau )}
\label{eq:62}
~~~.
\ee
It is straightforward to take the inverse Fourier transform of
Eq.(\ref{eq:61}) with the result
\be
f(12)=R(t_{1},t_{0})R(t_{2},t_{0})\left(\frac{\ell^{2}}{\tilde{L}^{2}}\right)
e^{-\frac{1}{2}\left(r/\tilde{L}\right)^{2}}
\label{eq:63}
\ee
where
\be
\tilde{L}^{2}=\ell^{2}+4b'T-2ib''(t_{1}-t_{2})
\label{eq:64}
\ee
and
\be
T=\frac{t_{1}+t_{2}}{2}
\label{eq:65}
~~~.
\ee
We must stop here and satisfy the constraint given by Eq.(\ref{eq:60}).
We have from Eq.(\ref{eq:63})
\be
1=R^{2}(t_{1},t_{0})\left(\frac{\ell^{2}}{{L}^{2}}\right)
\label{eq:66}
\ee
where
\be
L^{2}(t_{1})=\tilde{L}^{2}(t_{1},t_{1})=\ell^{2}+4b't_{1}
\label{eq:67}
~~~.
\ee
Eq.(\ref{eq:66}) serves as an equation for $\omega_{0}(t_{1})$ which
can be easily solved to give
\be
\omega_{0}(t_{1})=\frac{2}{L^{2}(t_{1})}=\frac{2}{\ell^{2}+4b't_{1}}
\label{eq:68}
~~~.
\ee
Using Eq.(\ref{eq:66}) to express $R(t_{1},t_{0})$ in terms of $L(t_{1})$
we find
\be
f(12)=\Phi (t_{1},t_{2})\frac{1}{1-i\omega}
e^{-\frac{1}{2}\left(r/\tilde{L}\right)^{2}}
\label{eq:69}
\ee
where
\be
\Phi (t_{1},t_{2})=\frac{L(t_{1})L(t_{2})}{L^{2}(T)}
\label{eq:70}
\ee
and
\be
\omega =\frac{2b''(t_{1}-t_{2})}{L^{2}(T)}
\label{eq:71}
~~~.
\ee
This last definition implies
\be
\tilde{L}^{2}=L^{2}(T)\left(1-i\omega \right)
\ee
and
\be
f(12)=\Phi (t_{1},t_{2})\frac{e^{-iz}}{1-i\omega}
e^{-\frac{1}{2}y^{2}}
\label{eq:72}
\ee
where
\be
y^{2}=\frac{x^{2}}{1+\omega^{2}}
\ee
\be
x=r/L(T)
\ee
and
\be
z=\frac{1}{2}\omega y^{2}
~~~.
\ee

There are a number of comments relevant to this result for $f(12)$
given by Eq.(\ref{eq:69}).  First note that there
is consistency between the definition
of $\omega_{0}(t)$ given by Eq.(\ref{eq:41}) and the solution for $f$
which leads to Eq.(\ref{eq:68}).
For equal times, $t_{1}=t_{2}=t$, we have
\be
f(r,t)=e^{-\frac{1}{2}x^{2}}
\label{eq:77}
\ee
which is of the same form as in the purely dissipative case\cite{EXP2,OJK}
with a characteristic
length $L\approx \sqrt{b't}$.  
It we look at the on-site $r=0$ autocorrelation function,
\be
f(0,t_{1},t_{2})=\Phi (t_{1},t_{2})\frac{1+i\omega}{1+\omega^{2}}
~~~,
\ee
we can write for $t_{1},t_{2}\gg t_{0}$,
\be
\Phi (t_{1},t_{2})=\left(\frac{\sqrt{t_{1}t_{2}}}{T}\right)^{\lambda_{0}}
~~~.
\ee
For $t_{1}\gg t_{2}$, $\omega$ approaches a constant and
the nonequilibrium exponent $\lambda_{0}$ for $\Phi$ also governs
$f(0,t_{1},t_{2})$ and  is given by
$\lambda_{0}=1$ which is the same\cite{EXP2}
as for the TDGL case for $n=d=2$.

The main new result is that for non-equal times the auxiliary
field correlation function shows a novel oscillatory behavior.
One of our chief goals below is discuss the possibility of
observing this phenomena.  We note here that $f(12)$ does obey a form
of scaling for $t_{1},t_{2}\gg t_{0}$:
\be
f(12)=f(x,\tau )=\Phi (\tau )\frac{e^{-iz(x,\tau)}}{1-i\omega (\tau )}
e^{-\frac{1}{2}\frac{x^{2}}{1+\omega^{2}(\tau )}}
\label{eq:79}
\ee
where $\tau =t_{1}/t_{2}$,
\be
\Phi (\tau )=\frac{2\sqrt{\tau}}{1+\tau}
\ee
\be
\omega (\tau )=\frac{b''}{b'}\left(\frac{\tau -1}{\tau +1}\right)
\label{eq:81}
\ee
and
\be
z(x,\tau)=\frac{1}{2}\frac{\omega (\tau )x^{2}}{1+\omega^{2}(\tau )}
~~~.
\label{eq:82}
\ee

Rather than discussing this result for $f(12)$ in more detail, it is prudent
to remember that $f(12)$ is not itself directly observable.  Thus
let us turn back to observables and their dependence on $f(12)$.
We delay discussing the details of the oscillations in $f(12)$ until
after discussing how this feeds back into the determination of observables.

\section{Defect-Defect Correlation Function}

\subsection{General Result}

Given the explicit solution for $f(12)$, Eq.(\ref{eq:79}), we can return 
to the evaluation of the density correlation function, $G_{\rho\rho}(12)$,
given by Eq.(\ref{eq:29}).  The input we need for its determination is 
$\gamma_{T}^{-2}=1-F^{2}$, where
\be
F^{2}=|f|^{2}=\frac{\Phi^{2}}{1+\omega^{2}}e^{-\frac{x^{2}}{1+\omega^{2}}}
\label{eq:83}
\ee
and
\be
Q=\left(f_{0}'\right)^{2}+\left(\Delta_{0}'\right)^{2} 
\nonumber
\ee
\be
=\frac{x^{2}F^{2}}{L^{2}\left(1+\omega^{2}\right)}
~~~.
\ee
Inserting these results for $\gamma_{T}$ and $Q$ back into Eq.(\ref{eq:29})
gives 
\be
G_{\rho\rho}(12)=\frac{\Phi^{2}(\tau )}{2\pi^{2} L^{4}(T)(1+\omega^{2}(\tau )}
g\left(\frac{x^{2}}{(1+\omega^{2}(\tau ))}\right)
\label{eq:87}
\ee
where
\be
g(s)=\frac{e^{s}(1-s)-\Phi^{2}(\tau )}{\left( e^{s}-\Phi^{2}(\tau )\right)^{2}}
~~~.
\ee
In analyzing $G_{\rho\rho}(12)$ we must be careful to distinguish the
equal-time case from the unequal times case.

\subsection{The equal-time case} 

If $t_{1}=t_{2}=t$ and $\tau =1$, the density correlation function
can be written as:
\be
G_{\rho\rho}(r,t)=\frac{1}{2\pi^{2} L^{4}(t)} g(x)
\ee
where
\be
g(x)=\frac{e^{x^{2}}(1-x^{2})-1}{\left( e^{x^{2}}-1\right)^{2}}
~~~.
\ee
This is the same result found in the purely dissipative case.  
It is known\cite{HAL81}
that the conservation of topological charge for
equal times 
requires one to include in the defect-defect correlation function the correlation
of a defect with itself:
\be
\tilde{G}_{\rho\rho}(r,t)=\delta ({\bf r}) \bar{n}(t)+G_{\rho\rho}(r,t)
\label{eq:91}
\ee
where $\bar{n}(t)$ is the average defect density.
Then conservation of topological charge is given by
\be
\int ~ d^{2}r ~ \tilde{G}_{\rho\rho}(r,t)=0 ~~~.
\label{eq:92}
\ee
Inserting Eq.(\ref{eq:91}) into Eq.(\ref{eq:92}) gives
\be
\bar{n}(t)=-\int ~ d^{2}r ~G_{\rho\rho}(r,t)
\label{eq:93}
~~~.
\ee
However using the form given by Eq.(\ref{eq:29}) we can do the
integral in Eq.(\ref{eq:93}) and obtain for the average defect density
\be
\bar{n}(t)=\lim_{r\rightarrow 0}\frac{Q\gamma_{T}^{2}}{2\pi}
=\frac{1}{2\pi L^{2}(t)}
~~~.
\ee
This is the expected  result if scaling holds.  One can also find 
$\bar{n}(t)$ by direct computation
and obtain
\be
\bar{n}(t)=\frac{\omega_{0} (t)}{4\pi}
\ee
where $\omega_{0}$ is defined by Eq.(\ref{eq:41}) and given  in this approximation
by Eq.(\ref{eq:68}). We see that the two
determinations of $\bar{n}(t)$ agree.

\subsection{The unequal-time case}

For the case $\tau \neq 1$ we have that the conservation of topological charge
holds directly for $G_{\rho\rho}(12)$ since
\be
\int ~ d^{2}r ~ G_{\rho\rho}(12) 
=-\lim_{r\rightarrow 0}\frac{Q\gamma_{T}^{2}}{2\pi} =0
~~~.
\ee
The final step follows 
since $Q\approx r^{2}$ for small $r$ and $\gamma_{T}^{2}$ is regular in this limit.
If we set $r=0$ in $G_{\rho\rho}(12)$ given by Eq.(\ref{eq:87}) we obtain
\be
G_{\rho\rho}(0,t_{1},t_{2}) =\frac{1}{2\pi^{2}L^{4}(T)}
\frac{1}{\left(1+\omega^{2}(\tau )\right)} \frac{4\tau}{(1-\tau )^{2}}
~~~.
\ee
We see that this quantity blows up a $\tau \rightarrow 1$ signaling the
existence of the $\delta$-function at $r=0$ obtained for equal times.  Thus
we see that the limits $r\rightarrow 0$ and $\tau \rightarrow 1$
do not commute.

\section{Defect Velocity Probability Distribution Function}

The 
defect velocity
probability distribution function is  defined by
\be
\bar{n}(t) P({\bf V})\equiv
\langle |\rho (\psi) | \delta ({\bf V}-{\bf v}[\psi])\rangle 
=\langle |\rho (m) | \delta ({\bf V}-{\bf v}[m])\rangle ~~~.
\ee
One of the main results from
the last section is that at equal times the auxiliary field probability
distribution is isotropic and has the same form as in the purely
dissipative case.  This means that we obtain the same result here
as found in Ref.(\onlinecite{N.8}) and given earlier by Eq.(\ref{eq:a1}) 
where the characteristic velocity $\bar{v}(t)$ 
is given by
\be
\bar{v}(t)=2\left(b'\right)^{2}\frac{S^{(4)}(t)}{\omega_{0}(t)}
\ee
where
\be
S^{(4)}(t)=\nabla^{4}f(r,t)|_{r=0}-\omega_{0}^{2}(t)
\ee
and using the explicit results for $f(r,t)$, given by Eq.(\ref{eq:77})
we obtain
\be
\bar{v}(t)=\frac{4\left(b'\right)^{2}}{L^{2}(t)}
~~~.
\ee

The result for $P({\bf V})$ given by Eq.(\ref{eq:a1}) indicates
that the probability of finding a defect with a large velocity decreases
with time.  However, since this distribution falls off only as
$V^{-4}$ for large $V$ only the first moment beyond the
normalization integral exists.  This seems to imply the existence of
a source of large velocities. Assuming the large velocities of defects
can be associated with the final collapse of a defect structure
(defect-antidefect pair annihilation for point defects), Bray\cite{bvvt}
used general scaling arguments to obtain the same
large velocity tail given by Eq.(\ref{eq:a1}).

One probe of the defect dynamics is to study the
correlation between two defects including the correlation between
their velocities.  The two-defect
velocity probability distribution, $P[{\bf V}_{1},{\bf V}_{2},{\bf r}]$,
gives the probability that two defects separated by a distance $r$
have velocities ${\bf V}_{1},{\bf V}_{2}$.  This quantity
was determined in Ref.(\onlinecite{vvv}) and since it is an equal time
quantity the results found there hold here. 
The  physical results from the calculation\cite{vvv} of this quantity for
the TDGL model, 
carried out in detail 
for $n=d=2$ using the same approximations as indicated above, are relatively 
simple to state.
For a given separation $r$ , the most probable 
configuration corresponds, as expected, to a state with zero total velocity and
a nonzero relative velocity only along the axis connecting the defects:
${\bf V}_{1}=-{\bf V}_{2}\equiv v\hat{x}$.
Moreover there is a definite most probable nonzero value 
for $v=v_{max}$ for a given 
value of $r$.  The most striking feature of these results is that for 
small $r$ the most probable velocity goes as $v_{max}=\kappa /r$ and
$\kappa =2.19$ in dimensionless units defined in Ref.(\onlinecite{vvv}).

\section{Order Parameter Correlation Function}

Thus far we have focussed on the statistical properties
of the defects in the system and found results for $G_{\rho\rho}$
and $P({\bf V})$ which are very similar to the purely dissipative
case.  Only rather small differences arise when one looks at
unequal times.  Unfortunately neither of these quantities
probes the full phase dependence of the auxiliary field correlation
function which shows interesting oscillations in space at
unequal times.  We show here that this phase dependence may be
probed via the order parameter correlation function evaluated
at unequal times.  Indeed this quantity, within the approximate
treatment given here, is
quite different from the purely dissipative case even at equal times.

The order parameter correlation function is defined by
\be
C_{\psi}(12)=<\psi^{*}(2)\psi (1)>
\label{eq:101}
\ee
and our approach toward its evaluation will be to
find the relationship between
the order parameter and the auxiliary field ${\bf m}$.  In section IV
we required that $\psi$ be proportional to ${\bf m}$ for small $m$
near the core of a defect.  In evaluating Eq.(\ref{eq:101}) we 
need a more general
mapping.  The procedure we will use here has been successful in 
the purely dissipative case\cite{TUG}.  Picking up on the point
made in section IV, we choose ${\bf m}({\bf x})$ to represent the distance 
from ${\bf x}$ to the closest defect.  This physical picture can
be realized by constructing $\psi ({\bf m})$ as a solution to
the equation for a single  stationary defect:
\be
b \nabla^{2}_{m}\psi +\left(1+i\omega_{1}-u|\psi |^{2}\right)\psi =0
~~~,
\label{eq:102}
\ee
where ${\bf m}$ serves as the coordinate and $\omega_{1}$ is
the oscillatory frequency.
In the purely dissipative case, $b$ real and $\omega_{1}=0$,
one has for field points well
away from a defect core:
\be
\psi ({\bf m}))=\psi_{0}e^{i\phi ({\bf m}) }
\label{eq:103}
\ee
where, for a defect of charge $n$,
\be
\phi ({\bf m}) = n ~ tan^{-1} \left( m_{y}  /m_{x}\right)
~~~.
\label{eq:104}
\ee
In the purely dissipative case, insertion of Eq.(\ref{eq:103}) for
$\psi ({\bf m})$ into Eq.(\ref{eq:101}) and carrying out the Gaussian average over
${\bf m}$, leads to the result\cite{Cont}:
\be
C_{\psi}(12)=\psi_{0}^{2}f\int_{0}^{1}dz\frac{\left(1-z^{2}\right)^{1/2}}
{\left(1-z^{2}f^{2}\right)^{1/2}}
\ee
with $f(12)$ given by Eq.(\ref{eq:69}) with $\omega =0$.  
This approximate result has been rather extensively tested in the
TDGL case\cite{BH}.

In the CGLE case we have a new and interesting
element.  There is a range of parameters where one has spiral defects.
Thus, unlike the TDGL case, one has spatial structure associated with
individual defects beyond the core. 
In particular Hagan\cite{Hagan} showed that the far-field solution 
of Eq.(\ref{eq:102})
is given by
\be
\psi ({\bf m}({\bf x}))=\psi_{0}e^{i\left(\phi ({\bf x})+q m({\bf x})\right) }
\label{eq:107}
\ee
where $q$ is the wavenumber of the spiral arms asymptotically far from
the defect core and $\psi_{0}=\psi_{0}(q)$.  $q$ depends on the particular 
parameters of the CGLE
as discussed by Hagan.  While there are values for which $q$ vanishes,
as in the TDGL limit, we will assume that we work in a region of 
parameter space where $q\neq 0$.

Using this set of mappings the order parameter correlation function is given by
\be
C_{\psi}(12)=\psi_{0}^{2}<e^{-i\left(\phi (2)+q m(2)\right) }
e^{i\left(\phi (1)+q m(1)\right) }>
\nonumber
\ee
\be
=\psi_{0}^{2}\frac{D^{2}}{(2\pi )^{2}}
\int ~ d^{2}x(1)d^{2}x(2)~e^{i(\phi_{1}-\phi_{2})}
~e^{iq(x(1)-x(2))}~e^{-\frac{1}{2}A}
\ee
where $d^{2}x(i)=x(i)dx(i)d\phi (i)$ for $i=1,2$.  The action $A$ and
determinant $D$ are given
in the Appendix. In particular $A$ is given by Eq.(\ref{eq:A13}) in terms 
of polar coordinates and we have
more explicitly,
\be
C_{\psi}(12)=\psi_{0}^{2}\frac{D^{2}}{(2\pi )^{2}}
\int_{0}^{\infty}x(1)dx(1)\int_{0}^{\infty}x(2)dx(2)
~e^{iq(x(1)-x(2))}~
\nonumber
\ee
\be
\times e^{-\frac{1}{2}\sum_{i}x^{2}(i)W_{0}(i)}
J(x(1),x(2))
\label{eq:109}
\ee
where the angular integrations are given by
\be
J(x(1),x(2))=\int_{0}^{2\pi} d\phi (1)\int_{0}^{2\pi} d\phi (2)
~e^{i(\phi (1)-\phi (2))}
~e^{D^{2}C_{T}x(1)x(2)~cos\left(\phi (1)-\phi (2)-\theta\right)}
\ee
and $\theta$ is defined by $tan ~\theta =\Delta /C_{0}$.  Shifting the angular
integrations we see that the $\theta$-dependence factors out:
\be
J(x(1),x(2))=2\pi ~e^{i\theta}\int_{0}^{2\pi} d\phi ~e^{i\phi } 
~e^{D^{2}C_{T}x(1)x(2)~cos \phi }
~~~.
\ee
If we change integration variables from $x(i)$ to
\be
y_{i}=\sqrt{W_{0}(i)}x(i)
~~~,
\label{eq:110}
\ee
we can rewrite Eq.(\ref{eq:109}) in the form
\be
C_{\psi}(12)=\psi_{0}^{2}\frac{f}{f_{T}}\gamma_{T}^{-2}
\int_{0}^{\infty}y_{1}dy_{1}\int_{0}^{\infty}y_{2}dy_{2}
~e^{-\frac{1}{2}\left(y_{1}^{2}+y_{2}^{2}\right)}
e^{i(q_{1}y_{1}-q_{2}y_{2})}
\frac{1}{2\pi}\int_{0}^{2\pi} d\phi ~e^{i\phi }
~e^{f_{T}y_{1}y_{2}~cos \phi }
\label{eq:111}
\ee
where
\be
q_{i}=q\sqrt{S_{0}(i)}\gamma_{T}^{-1}
~~~.
\label{eq:112}
\ee
Notice that the phase dependence of the auxiliary field correlation
function  is isolated in the
overall factor of $f$ in Eq.(\ref{eq:111}).
The integral over $\phi$ in Eq.(\ref{eq:111}) gives a modified
Bessel function, but for our purposes we only need the power-series
result:
\be
\frac{1}{2\pi}\int_{0}^{2\pi} d\phi ~e^{i\phi }
~e^{f_{T}y_{1}y_{2}~cos \phi }
=\sum_{k=0}^{\infty}\frac{1}{k!(k+1)!}
\left(\frac{f_{T}y_{1}y_{2}}{2}\right)^{2k+1}
\ee
and
\be
C_{\psi}(12)=\psi_{0}^{2}\frac{f}{2}\gamma_{T}^{-2}
\sum_{k=0}^{\infty}\frac{1}{k!(k+1)!}\left(\frac{f_{T}}{2}\right)^{2k}
J_{k}(q_{1})J_{k}^{*}(q_{2})
\label{eq:113}
\ee
where
\be
J_{k}(q_{1})=\int_{0}^{\infty}~ydy~e^{-\frac{1}{2}y^{2}}e^{iq_{1}y}
y^{2k+1}
~~~.
\label{eq:114}
\ee
In the limit $q\rightarrow 0$ Eq.(\ref{eq:113}) does, after some
manipulations, reduces to the result found in the TDGL case for $n=2$.
Note that for $q\neq 0$, except for $r=|{\bf r}_{1}-{\bf r}_{2}|$
very small,  $q_{i}$, given by Eq.(\ref{eq:112}), is becoming
increasingly large with $\sqrt{S_{0}(i)}\approx L(t_{i})$.  This
means we need evaluate $J_{k}(q_{i})$ only for large $q_{i}$.  
Evaluation of  $J_{k}(q_{i})$ for large $q_{i}$ is facilitated by
writing 
\be
J_{k}(q_{i},a)=\int_{0}^{\infty}~ydy~e^{-a y^{2}}e^{iq_{i}y}
y^{2k+1}
\nonumber
~~~,
\ee
where $J_{k}(q_{i})=J_{k}(q_{i},1/2)$.  We have then
\be
J_{k}(q_{i},a)
=\left(-\frac{\partial}{\partial a}\right)^{k+1}J(q_{i},a)
\label{eq:116}
\ee
and
\be
J(q_{i},a)=\int_{0}^{\infty}dy~e^{-a y^{2}}e^{iq_{i}y}
\nonumber
\ee
\be
=\sqrt{\frac{\pi}{2a}}e^{-\frac{q_{i}^{2}}{4a}}+iJ''(q_{i},a)
~~~.
\ee
We see that the real part of $J$ is exponentially small
for large $q_{i}^{2}$. However, it is easy to see that for large
$q_{i}$
\be
J''(q_{i},a)=\frac{1}{q_{i}}+\frac{2a}{q_{i}^{3}}
+\frac{3(2a)^{2}}{q_{i}^{5}}+\cdots
\ee
This means that the leading non-exponential contribution to the 
order-parameter correlation function comes from $J_{0}(q_{1})$ and is
given to leading order by
\be
J_{0}(q_{1},a)
=\left(-\frac{\partial}{\partial a}\right)i\frac{2a}{q_{i}^{3}}
+\cdots =-\frac{2i}{q_{i}^{3}}+\cdots
~~~.
\ee
Inserting this result back into Eq.(\ref{eq:113}) we obtain
\be
C_{\psi}(12)=\psi_{0}^{2}\frac{f}{2}\gamma_{T}^{-2}
J_{0}(q_{1})J_{0}^{*}(q_{2})
\nonumber
\ee
\be
=\psi_{0}^{2}\frac{f}{2}\gamma_{T}^{-2}\frac{4}{(q_{1}q_{2})^{3}}
\nonumber
\ee
\be
=\frac{\psi_{0}^{2}}{q^{6}} 
\frac{2f\gamma_{T}^{4} }{(S_{0}(1)S_{0}(2))^{3/2}}
~~~.
\label{eq:120}
\ee
The scaled portion of the OP correlation function for $x\neq 0$
can be written as
\be
W(x,\tau )=\frac{q^{6}}{\psi_{0}^{2}}S_{0}^{3/2}(t_{1})
S_{0}^{3/2}(t_{2})C_{\psi}(12)
\nonumber
\ee
\be
=2f\gamma_{T}^{4}
~~~.
\ee
For $\tau \neq 1$, the correlation function depends strongly on
$\eta =b''/b'$ via $\omega$.  This is not true at equal times
where $\eta =b''/b'$ does not appear in the scaling function.
For $\tau \neq 1$ the oscillations in $f$ are now clear in the
order parameter correlation function.
Writing out the real and imaginary parts we obtain
\be
W'=\frac{2F}{\sqrt{1+\omega^{2}}}\frac{1}{\left(1-F^{2}\right)^{2}}
\left[cos~z +\omega sin~z\right]
\ee
\be
W''=\frac{2F}{\sqrt{1+\omega^{2}}}\frac{1}{\left(1-F^{2}\right)^{2}}
\left[-sin~z +\omega cos~z\right]
~~~,
\ee
where $F$ is given by Eq.(\ref{eq:83}) and $\omega$ and
$z$ by Eqs.(\ref{eq:81}) and (\ref{eq:82}).
We are interested in the oscillations associated with 
$\eta =b''/b' \neq 0$.  These are most clearly manifested in
$W''$ and characterized by the zeros at
\be
\omega = tan ~ z_{0}
\ee
The first zero as a function of scaled distance is given by
\be
x_{0}^{2}=2+\frac{4}{3}\omega^{2}+\cdots
\ee
for small $\omega$ and
\be
x_{0}^{2}=\pi\omega -2 +{\cal }(1/\omega)
\ee
for large $\omega$.

\section{Conclusions}

By using some new ideas about how to characterize defect dynamics,
we have
shown how one can determine local expressions for the defect density
and defect velocity in terms of derivatives of the order parameter
fields.  These exact results were then used to derive approximate results
for the defect-defect density correlation function, defect velocity
probability distribution, and the order parameter correlation functions.
Within these approximations, which work well for the purely
dissipative case, we find that the results for the defect-defect 
density correlation function and the defect velocity 
probability distribution are substantially unchanged from the
TDGL case.  Thus these results seem robust.  The results
for the two-time auxiliary field correlation function indicate some
interesting oscillation of its phase as a function of scaled
distance.  Since the defect-defect density correlation function
depends only on the amplitude of the auxiliary field correlation function
these oscillations are not present.  In the last section we
have seen that some remnant of these oscillations is present in
the order parameter correlation function.  However, another new element
element for the order parameter correlation function  is that the spiral 
arms for the defects render
the interactions between different spatial points much shorter range
than for the purel dissipative case.  Thus for different spatial points
at equal times the order parameter correlation function  is down by a factor 
of $\bar{n}^{3}(t)$ relative
to the TDGL case.  All of these results can be tested via  numerical
simulation.

Acknowledgement:  This work was supported in part by the MRSEC
program of the National Science Foundation under Contract No.
DMR-9808595.  I thank Drs. I. Aranson and P. Kevrekidis for very
useful discussions.

\appendix
\section{}

In the purely dissipative case all correlations are isotropic:
\be
C_{\nu\nu'}(ij)=<m_{\nu}(i)m_{\nu'}(j)> =\delta_{\nu\nu'}C_{0}(ij)
\ee
where $(i,j)=(1,2)$.  In the complex case, over time, the real and
imaginary components of the order parameter are mixed, and this
requires that we treat the more general correlation function for the auxiliary
field
\be
C_{\nu\nu'}(ij)=\delta_{\nu\nu'}C_{0}(ij)
+\epsilon_{\nu\nu'}\epsilon_{ij}\Delta (12)
\label{eq:A2}
\ee
which satisfies the required symmetry for classical fields,
\be
C_{\nu\nu'}(ij)=C_{\nu'\nu}(ji)
\ee
if $C_{0}(ij)=C_{0}(ji)$.  Thus the variance of the gaussian field
${\bf m}$ is determined by the two independent functions
 $C_{0}(12)$ and $\Delta (12)$.

We will be concerned with various two-point averages over ${\bf m}$
of the general form:
\be
C_{AB}(12)=<A({\bf m}(1))B({\bf m}(2))>
=\int~d^{2}x(1)~d^{2}x(2) A({\bf x}(1))B({\bf x}(2))
\Phi ({\bf x}(1),({\bf x}(2))
\ee
where the two-point probability distribution is given by
\be
\Phi ({\bf x}(1),({\bf x}(2))
=<\delta ({\bf x}(1)-{\bf m}(1))\delta ({\bf x}(2)-{\bf m}(2))>
\nonumber
\ee
\be
=\int \frac{d^{2}k(1)}{(2\pi)^{2}}\frac{d^{2}k(2)}{(2\pi)^{2}}
e^{i\sum_{j}{\bf k}(j)\cdot{\bf x}(j)}
exp.\left(-\frac{1}{2}\sum_{\nu\nu'}\sum_{ij}
k_{\nu}(i)k_{\nu'}(j) C_{\nu\nu'}(ij)\right)
\nonumber
\ee
\be
=\frac{1}{(2\pi)^{2}}\frac{1}{(det ~C)^{1/2}}
e^{-\frac{1}{2}A}
\ee
where the argument of the exponential is given by
\be
A=\sum_{\nu\nu'}\sum_{ij}
x_{\nu}(i)x_{\nu'}(j) W_{\nu\nu'}(ij)
\ee
and the matrix $W$ is the inverse of $C$ defined by
\be
\sum_{\gamma k}W_{\nu\gamma}(ik)C_{\gamma\nu'}(kj)=\delta_{\nu\nu'}\delta_{ij}
~~~.
\ee
$W$ is given explicitly by
\be
W_{\nu\nu'}(ij)=\delta_{\nu\nu'}D^{-2}
\left[ \delta_{ij}\left(\frac{S_{0}(1)S_{0}(2)}{S_{0}(i)} +C_{0}\right)
-C_{0}\right]-\epsilon_{\nu\nu'}\epsilon_{ij}D^{-2}\Delta
\ee
where
\be
S_{0}(i)=C_{0}(ii)
\ee
\be
D^{-2}=S_{0}(1)S_{0}(2)-C_{T}^{2}
\label{eq:A10}
\ee
\be
C_{T}^{2}=C_{0}^{2}+\Delta^{2}
\ee
and finally
\be
det ~ C=D^{-4}
~~~.
\ee

If we express ${\bf x}(i)=x_{i}(cos~\phi_{i},sin~\phi_{i})$, the argument of the
exponential in the distribution takes on the simple form
\be
A=\sum_{i}x_{i}^{2}W_{0}(i)-2D^{-2}C_{T}x_{1}x_{2}cos(\phi_{1}-\phi_{2}-\theta )
\label{eq:A13}
\ee
where
\be
C_{0}=C_{T}~ cos~\theta
\label{eq:A14}
\ee
\be
\Delta=C_{T}~ sin~\theta
\label{eq:A15}
\ee
and
\be
W_{0}(i)=D^{-2}\frac{S_{0}(1)S_{0}(2)}{S_{0}(i)}
=S_{0}^{-1}(i)\gamma_{T}^{-2}
\label{eq:A16}
~~~.
\ee


\begin{thebibliography}{99}
\bibitem{gauge}  It turns out, as discussed carefully in
Ref.(\onlinecite{Tornkvist}), that one can replace $\theta_{B}$ and
$w$ by $\theta$ and $ln R$ in Eq.(\ref{eq:40})
\end{thebibliography}
\end{document}